\documentclass[12pt,preprint]{aastex}

\newcommand{\iso}[2]{\hbox{${}^{#1}{\rm #2}$}}
\newcommand{\Msun}{\ensuremath{{\rm M}_{\sun}}}

\shorttitle{The Chemical Evolution of Helium in GCs}
\shortauthors{Karakas et al.}

\begin{document}


\title{The Chemical Evolution of Helium in Globular Clusters: \\
  Implications for the Self-Pollution Scenario}


\author{A. I. Karakas\altaffilmark{1}}
\affil{Department of Physics \& Astronomy, McMaster University, 1280 Main
     Street W., Hamilton ON L8S 4M1, Canada}

\author{Y. Fenner}
\affil{Harvard-Smithsonian Center for Astrophysics, 60 Garden Street, 
     Cambridge MA 02138 USA}
\email{yfenner@cfa.harvard.edu}

\author{Alison Sills}
\affil{Department of Physics \& Astronomy, McMaster University, 1280 Main
     Street W., Hamilton ON L8S 4M1, Canada}
\email{asills@mcmaster.ca}

%\and

\author{S. W. Campbell and J. C. Lattanzio}
\affil{Centre for Stellar \& Planetary Astrophysics, School of Mathematical Sciences,
  PO Box 28M, Monash University, Clayton VIC 3800, Australia}
\email{Simon.Campbell@maths.monash.edu.au, John.Lattanzio@sci.monash.edu.au}


\altaffiltext{1}{H. G. Thode Fellow at the Origins Institute}


\begin{abstract}
We investigate the suggestion
that there are stellar populations in some globular clusters 
with enhanced helium ($Y \sim$ 0.28 to 0.40)
compared to the primordial value.  We assume that a previous
generation of massive Asymptotic Giant Branch (AGB) stars have polluted
the cluster. Two independent sets of AGB yields are used to follow 
the evolution of helium and CNO
using a Salpeter initial mass function (IMF) and two top-heavy
IMFs.  In no case are we able to produce the postulated
large $Y \sim 0.35$ without violating the observational
constraint that the CNO content is nearly constant.
\end{abstract}




\section{Introduction}

\par
Star-to-star abundance variations of the light
elements C, N, O, Na, Mg and Al have been observed in every well studied globular
cluster (GC) to date \citep[][and references therein]{kraft94,gratton04}
but are not found in field stars of the same metallicity \citep{gratton00}.
Hence these abundance anomalies are somehow the result of the cluster
environment. The variations of the elements follow a common pattern:
C-N,  O-Na and Mg-Al are all
anti-correlated \citep{shetrone96a,kraft97,cannon98,gratton01,cohen05a,cohen05b}.
The abundances of iron-peak, s and r-process elements do not show the
same star-to-star scatter nor do they vary {\em with} the light
elements \citep{gratton04,james04,yong06b}, although new observations
by \citet{wylie06} suggest that there is real star-to-star scatter
amongst heavy element abundances in stars in the metal-rich cluster 47 Tucanae.
The other important exception is the massive cluster $\omega$ Centauri, whose 
age and metallicity spread, along with
a rise in s-element abundance with increasing [Fe/H], suggests that
it evolved very differently from other GCs and may even
have an extragalactic origin \citep{smith00}.
The key points are that O has been destroyed in some stars by up to one
dex, the C$+$N$+$O abundances remain almost constant and there is no evidence
for large-scale variation of the neutron-capture elements.
\par
Two hypotheses had been proposed to explain these observed abundances.
The first was deep mixing, where the abundance
anomalies are produced by internal mixing
during the ascent of the giant branch, after the first dredge-up
\citep{sweigart79,pin97,charbonnel94}.
However, star-to-star abundance variations in C, N, O and Na were
subsequently observed in stars at or near the main-sequence turn-off
\citep{gratton01,ramirez03,james04,cohen05a,cohen05b}.
These observations support the self-pollution scenario,
first proposed by \citet{cottrell81}. Here
a previous generation of stars polluted the atmospheres of stars we
observe today or provided part of the material from which those
stars formed.
Because [Fe/H] is roughly  constant in stars in a given GC it has
been assumed that polluters were intermediate-mass
AGB stars with initial masses between
$\sim$3 to 8$\Msun$ rather than supernovae, which produce Fe.
The hot bottom burning experienced by these stars provides the
hydrogen burning environment (at least qualitatively) 
 that can alter the abundances of the light elements.
One consequence is the production of a significant quantity
of helium \citep[$Y$, \iso{4}He;][]{lattanzio04}.
The mass lost via the slow winds of AGB stars could, in principle,
have been retained by the cluster from which new stars may have
been born \citep{thoul02}.
Detailed AGB models have so far mostly failed to match the observed
abundance trends \citep{herwig03b,fenner04,campbell04,cohen05b},
but major uncertainties that affect the models undermine the
reliability of the predictions and leave room for an AGB
solution \citep{ventura05a,ventura05b}.
\par
Horizontal branch (HB) and main-sequence color-magnitude diagrams (CMD)
\citep{norris04,dantona04,lee05,piotto05} provide increasing evidence 
for helium enrichment in some GC stars.
%This is also proposed  to have come from a previous generation
%of low-metallicity intermediate-mass AGB stars. 
The unusual HB morphology of NGC 2808, which exhibits an extended
blue tail and a gap separating the red and blue clumps \citep{bedin00},
can be most easily explained if the blue stars have a higher helium content
($Y \sim 0.32$) compared to those in the red clump which 
presumably have primordial $Y \approx 0.24$ \citep{dantona04}.
Furthermore, NGC 2808 also has a peculiar main sequence 
\citep{dantona05} where the bluer stars are inferred to have
$Y \sim 0.4$ from fitting theoretical isochrones to the observed
data.  To reproduce the CMD, \citet{dantona05} note the absolute 
necessity of including a small population ($\sim 20$\%) of stars 
with $Y = 0.40$ and assume a spread of $Y$ between 0.24 -- 0.29 
to fit the main fraction ($\sim 80$\%).  While the 
exact maximum value of $Y$ required to reproduce the CMD is dependent 
on the color-$T_{\rm eff}$ transformations used in the analysis and 
therefore rather uncertain, it seems that the most plausible explanation
is that the bluest stars have enhanced amounts of helium with 
$Y \gtrsim 0.35$.  Observations by \citet{piotto05} show that
the blue main-sequence of $\omega$ Centauri is more metal-rich than the
red sequence, contrary to what is expected from stellar models
and \citet{norris04} showed that isochrones with $Y=0.40$ best
fit the bluest stars.  Until a better explanation for these
intriguing observations is found, we take them as motivation to 
study the AGB self-pollution scenario from a global perspective.
\par
In this paper, we use the \citet{fenner04} GC chemical evolution model to
follow the evolution of helium in the intracluster gas. We
probe AGB model uncertainties by using two independent sets
of yields, including those used in the previous study which were
tailor made for NGC 6752 (with a metallicity
[Fe/H] $\approx -1.4$\footnote{Slightly more metal-rich than the average
of [Fe/H] $\sim -1.6$ for NGC 6752, M13, M3 and
$\omega$ Cen with metallicities from \citet{harris96}; NGC 2808 has
[Fe/H] $\sim -1.15$.}).
The evolution of C, N and O is also followed since they impose
important empirical constraints, i.e. C$+$N$+$O $\approx$ constant,
that must be met by the model.

\section{Helium Production in AGB Stars} \label{sec:heprod}

Prior to the AGB both the first and second dredge-up (SDU)
mix helium to the surface from regions that have undergone some
H burning. During the thermally-pulsing-AGB phase,
partial He-burning results in an abundance of $Y \sim 0.75$ in the
intershell region, and each third dredge-up (TDU) episode
increases the \iso{4}He abundance of the envelope.
Hot bottom burning (HBB) also produces \iso{4}He via the CNO cycle.
For massive AGB stars the most important mixing phase is the SDU
which results in an increase of $\Delta Y \lesssim 0.1$.  The first
dredge-up is inefficient in low-$Z$ stars over 3$\Msun$
(see Fig.~2 in \citet{boothroyd99}), and efficient TDU and HBB result
in small increases of at most $\Delta Y \approx 0.03$, depending on the
time spent on the AGB and the temperature at the base of the convective
envelope.
\par
The helium yields from the AGB models with $Z=0.0017$ 
used by \citet{fenner04}
are shown in Table~\ref{table1} as the average mass fraction of 
$Y$ in the winds and the total mass of helium expelled into the
intracluster medium by each model. We hereafter refer to the 
Campbell et al. models as ``our'' models. 
We also show the $Z=0.001$ yields from \citet{ventura02} for 
comparison and note that the yields agree to within $\sim 30$\%.
Our yields (and average $Y$) are 
systematically larger for $m > 3.5\Msun$, reflecting the 
different input physics used in the two computations. 
\citet{ventura02} use a different convective model (Full Spectrum of
Turbulence instead of the mixing-length theory) and 
mass-loss rate, and observe shallower dredge-up.  
In Table~\ref{table1} an important result can be seen -- 
the average $Y$ from our models does not monotonically 
increase with increasing stellar mass but instead peaks at 5$\Msun$. 
%possibly because of dilution with a bigger Menv? and deeper TDU in the 5?
We observe efficient TDU and HBB plus our models spend longer
on the AGB thanks to the \citet{vw93} mass-loss rate
\citep[see discussion in][]{karakas06a}.
On the other hand, the models of \citet{ventura02} have more
efficient envelope convection resulting in larger luminosities
and shorter AGB lifetimes, owing to their choice of
a luminosity-driven mass-loss rate \citep{ventura05a}.  
This results in smaller helium yields and consequently less 
helium in the cluster gas for the Ventura et al. yields.
This difference has important consequences for the chemical
evolution model, discussed further in \S\ref{sec:results}.
\par
Previously \citep{karakasThesis} we compared the
stellar yields from the Monash models with those from
\citet{forestini97,vandenhoek97,marigo01} and \citet{izzard04},
for varying metallicities and find agreement for \iso{4}He for
$m \ge 5\Msun$ at the level of $\sim 30$\%, with the exception of
\citet{vandenhoek97} who produce $\sim 90$\% less \iso{4}He.
\citet{karakasThesis} also observed that the final surface abundance 
of $Y$ in the $Z=0.004$ models (slightly higher but similar 
to the average $Y$ in the winds) did not monotonically increase 
with mass but peaked at both $2.5\Msun$ (owing to efficient TDU) 
and $6\Msun$ (due to hot bottom burning).
We conclude that the relatively close agreement between
helium yields from various studies indicates they are more robust
than other species (e.g. \iso{12}C and \iso{16}O). This is partly
because the net result of hydrogen fusion is
helium production regardless of the rates of the various internal
cycles (CNO cycle, NeNa and MgAl chains).

\section{The Chemical Evolution Model} \label{sec:model}

The GC chemical evolution (GCCE) model was 
described in detail in \citet{fenner04}; here we
summarize the main features and the changes made for this study.
We assume a two-stage formation model whereby the first stage acts
as a prompt initial enrichment  that brings the cluster gas 
up to a metallicity of [Fe/H] = $-1.4$.  This first stage assumes 
a bimodal top-heavy IMF \citep{nakamura01} and zero-metallicity
massive star yields from  \citet{chieffi02} and \citet{umeda02}.
\par
During the second stage, we assume that the GC stars form in
10$^{7}$ years from this low-metallicity gas\footnote{The models
of \citet{ventura02} used scaled-solar abundances
whereas those of \citet{fenner04} used initial abundances
taken from the prompt initial enrichment stage. This also
accounts for the larger than expected $Z=0.0017$ compared to
$Z \approx 0.001$ that we would get if we assumed scaled-solar initial
abundances.
We therefore shift the [O/Fe] values from \citet{ventura02}
by a factor of 2.5 to account for this difference.}.
In \citet{fenner04}, the \citet{kroupa93} IMF was adopted and
it was assumed that the GC retained ejecta from stars with
$m \le 6.5\Msun$; here we change to a standard Salpeter-like
IMF \citep{salpeter} and test the effect of different  power-law slopes
(see \S\ref{sec:imf}).  Furthermore we also run separate simulations
using the
AGB yields from \citet{ventura02} for helium and the CNO isotopes.
These yields cover the mass range $3 \le m (\Msun) \le 5.5$;
we extrapolate these yields to 2.5$\Msun$ and 6.5$\Msun$ and
substitute in yields from \citet{fenner04} for $m < 2.5\Msun$.
We underline here that this extrapolation to higher mass renders 
uncertain the results obtained in terms of the maximum $Y$ 
expected after 50 $-$ 100 Myr when using the \citet{ventura02} 
yields. Although, in this case the extrapolation should be 
fairly reasonable given that the yields are monotonic with mass.
No contribution from Type Ia SNe was included due
to the observed uniform  [Fe/H].
We favor a scenario in which  the next (third) generation of stars 
is assumed to {\em form} out of this polluted gas rather 
than this gas accreting onto their surfaces.
This is because observational evidence shows that there is no
dilution of the surface abundances when stars move through
the first dredge-up.  The first dredge-up would mix the polluted
envelope material with primordial material, yet no changes to 
O, Na, Mg or Al abundances are observed at this stage
\citep{gratton01}.
\par
In the GCCE model we simply track the 
composition of the cluster gas as a function of time. 
Hence the abundance of the gas reflects the continuous 
addition of AGB material as  stars of decreasing mass 
evolve and lose their envelopes via winds.  
We do not model the formation of the third stellar generation 
but we can speculate that star formation will lock
up some of this gas in new stars at a rate (and efficiency) 
that  does not use up all of the gas from the 6.5$\Msun$ 
stars before the 6$\Msun$ stars have added new material 
($\sim 20$ Myr). This is going on star formation timescales
for low-mass stars, which can be anything from $\sim 1$ Myr
during the proto-star phase to $\sim 10^{8}$ Myr to reach the 
zero-aged main sequence \citep{siess00,white04}.

\subsection{The Initial Mass Function}  \label{sec:imf}

The IMF is one of the most uncertain parameters of the GCCE model.
Hints that the IMF is not universal \citep{kroupa01} have come from 
observations of carbon-enhanced s-process-rich metal-poor stars 
\citep{lucatello05}, which suggest more s-process-producing AGB stars 
were required at the earliest stages of Galactic formation.
Observations of the present-day mass function for
the Arches cluster in the Galactic center by \citet{stolte05} suggest a 
turnover mass of $\sim 6\Msun$, with a possible absence of low-mass stars. 
\citet{dantona04} required a factor of $\sim 10$ more  4 -- 7$\Msun$ 
stars than produced by a Salpeter-like IMF to return the amount of
helium required to form the number of blue HB stars in NGC 2808.
The lack of observational evidence for enhanced levels of s-process 
elements or carbon suggests that stars between 1 -- 3$\Msun$ are 
either not produced in significant numbers or they are ejected from 
the cluster. 

Dynamical simulations of GCs suggest that top-heavy IMFs result in
larger velocity dispersions than when using a Salpeter-like
IMF (Downing \& Sills, in preparation), and that cluster masses grow
too large if the IMF is normalized by the number of blue HB stars.
Simple analytical considerations led \citet{bekki06} 
to conclude that the top-heavy IMFs required to produce the postulated 
helium enrichment in $\omega$ Cen would most likely result in the 
disintegration of the cluster.  With these points in mind, we 
compute separate simulations with 1) a Salpeter-like IMF and 2) 
a top-heavy IMF 
that increases the number of intermediate-mass AGBs by a factor of 10;
see Figure~\ref{fig1}.  We hereafter refer to the 
top-heavy IMF as ``IMS-enhanced'' (intermediate-mass star enhanced).
In \citet{karakas06b} we present results from simulations that
employ a power-law IMF with slope of 0.3.

\section{Results}  \label{sec:results}

Previously,  \citet{fenner04} conservatively 
assumed that at the end of the simulation the intracluster gas 
consists of about three parts AGB ejecta to one
part primordial material.  If we think about this assumption in
terms of $Y$, then we begin with an initial $Y = 0.23$ and
after about 50 Myr the most massive AGB stars of $\sim 6.5\Msun$
start losing mass and hence we see an 
increase in $Y$ to some value dictated by the maximum
$Y$ in the AGB ejecta and the amount of dilution with
primordial gas.  The shape of the IMF will also dictate when the
$Y$ peak is reached, where a standard Salpeter IMF will favor
the contribution from lower mass AGB stars that produce significant
helium. The peak will be offset in time according to the lifetime
of these stars ($\sim 4 - 5\Msun$) and this is indeed the 
behavior we see in Figure ~\ref{fig2} (a), where we show the 
time evolution of helium using the standard Salpeter IMF 
for both sets of yields.  In the case with some dilution 
with primordial material (the top panel {\em a}) the 
helium abundance is not predicted to exceed $Y > 0.30$ at any 
time, where our models predict a maximum $Y \approx 0.29$ 
and Ventura et al.'s $Y \approx 0.26$, both lower than 
the inferred maximum helium ($Y \gtrsim 0.30$) in HB 
and blue main-sequence stars.
In  \S\ref{sec:heprod} we described that the helium 
abundance in our AGB models does not monotonically increase 
with stellar mass, but instead peaks at $\sim 5 \Msun$. 
On the other hand, the \citet{ventura02} yields are monotonic 
with mass so we expect a maximum $Y$ at the slightly earlier
times than when using our yields.  From Figure~\ref{fig2} we
see that the maximum $Y$ is reached in just over $\sim 100$ Myr,
reflecting the dominant contribution of stars with $m \sim 4 - 5\Msun$, 
favored both by the Salpeter IMF and by the amount of 
helium they produce.
The difference in maximum $Y$ between the sets of yields can be
attributed to the different input physics as discussed in 
\S\ref{sec:heprod}.
\par
One of the consequences of deep TDU is the production of \iso{12}C
which is quickly converted to \iso{14}N by HBB at the
base of the convective envelope. In Figure~\ref{fig3} and~\ref{fig4} 
we show the temporal evolution of C, N, and O (on a $\log$ scale) as
a function of the helium mass fraction.
Using a standard Salpeter IMF (Figure~\ref{fig3}) we see that 
the deep TDU observed in our models result in a $\sim 0.8\,$dex 
increase in C$+$N$+$O with most of this in the form of \iso{14}N,
again reflecting the dominant contribution of intermediate-mass
AGB stars. The Ventura et al. yields show a
modest increase in C$+$N$+$O of only $\sim 0.4$~dex and
oxygen has been destroyed by 0.3~dex in comparison to
our results where we see little or no O depletion.
The large increase in CNO that the Campbell et al. yields 
predict is not observed
in GC stars whereas the small increase from the Ventura et al.
yields is probably within observational uncertainties.

The results of the simulation using the IMS-enhanced IMF are
shown in Figure~\ref{fig4}. Using our yields, we
see a substantial increase in helium, $Y \approx 0.35$,
similar to the value required by isochrones \citep{dantona04,dantona05}
to match the bluest HB and main-sequence stars of NGC 2808
and $\omega$ Centauri. This increase is also accompanied by
a 1~dex increase in CNO although now there is a noticeable
depletion of O by $\sim 0.3$~dex.
The simulation using Ventura et al. yields maintains a constant
CNO (to within $\sim 0.3$~dex) but in this case the maximum $Y$
does not exceed 0.30. Importantly, the substantial O depletion
of $\sim 0.8$ dex in this case is similar to the maximum dispersion 
observed in GC stars (ignoring the case of the peculiar M13, 
where the most O depleted stars with [O/Fe] $\sim -1$ are 
likely the result of enhanced extra-mixing).  If we compare
with observations, the most ``polluted'' stars in a number of 
different clusters including M5, M15, M71 and NGC 6752 have
[O/Fe] $\approx -0.5$ \citep{ramirez02} whereas ``normal'' stars
have [O/Fe] $+0.5$, indicating significant O destruction of
about 0.8 -- 1 dex.  The simulation with the
IMS-enhanced IMF gives a higher weight to the most massive
AGB stars which tend to destroy O at the base of the envelope at
temperatures $T \gtrsim 80 \times 10^{6}$K.
\par
The dilution with primordial material only affects 
the smoothness of the resulting curve ($Y$ as a
function of time) and the maximum $Y$ in the cluster gas.
To check this is the case we compute two simulations with
pure AGB ejecta and no dilution at all, one for each set 
of yields. We show the results of these models in
Figure~\ref{fig2} (b), where we show the evolution of $Y$, 
and in Figure~\ref{fig5}, where we show evolution of the 
CNO species as a function of $Y$. The behavior of $Y$ 
with mass is seen more clearly in these figures, where
the simulation using our yields peaks at $Y \approx 0.36$, 
similar to the maximum $Y$ in the 5$\Msun$ model. 
Note also that $Y$ keeps increasing after the first 
initial jump from 0.23 to 0.34.
The simulations using Ventura et al. yields peak very sharply
at $Y \approx 0.30$  before decreasing smoothly with time.
The behavior of  C$+$N$+$O  is most interesting
here.  Our models  predict a significant increase of 
$\sim 1$ dex,  as in the previous simulations with some dilution.
The simulation using the Ventura et al. yields now also
show a large spread in C$+$N$+$O, varying by $\sim 0.8$ dex.
This is because the diluted primordial gas is essentially
nitrogen-free, off-setting increases from HBB.

\section{Discussion}  \label{sec:discussion}

From the results presented in the previous section, it seems
clear that the AGB self-pollution scenario does not successfully
match the observational constraints that we have considered
in this study, namely that C$+$N$+$O  is roughly constant and
that the maximum $Y \gtrsim 0.30$. In no case, regardless of
yields or the shape of the IMF, were we able to 
simultaneously match these constraints. 
The maximum helium abundances inferred from 
theoretical modeling \citep{dantona04,dantona05}
of the CMDs are quite uncertain and we should give more weight
to the spectroscopic observations 
showing C$+$N$+$O $\approx$ constant. We briefly consider the
simulations that manage to fit this constraint.

The simulations that most closely match the observed CNO 
data are those  employing the Ventura et al.
yields, regardless of the shape of the IMF (although see
\citet{karakas06b} for models with a flat power-law), but with
a small amount of dilution of primordial gas. Note that 
some dilution is  likely, since the star 
formation efficiency of low-mass stars is quite small, of 
the order of $\lesssim 50$\% \citep{matzner00}. 
Even assuming a top-heavy IMF such as the IMS-enhanced
IMF used in our study, a low-star formation efficiency 
and {\em no} primordial material would make the job of producing
enough polluted stars challenging.

There are many stellar model uncertainties:  
In particular, the extent of the TDU is far from known 
and shallower dredge-up, as observed in the 
Ventura et al. models, would help keep C$+$N$+$O constant while 
moderating the maximum $Y$.  If this was combined with a long 
enough HBB lifetime, implying low AGB mass-loss rates such as 
those obtained when using the \citet{vw93} prescription, then 
the required abundance patterns may be obtained by the essentially 
pure HBB environment. The HBB lifetime is also dependent on the 
convective model, and as shown by \citet{ventura05a} more efficient
convection, coupled with a luminosity-driven mass-loss
rate, results in a shorter AGB lifetime.

If abundances
of the bluest stars are closer to $Y \sim 0.3$, instead
of $Y \sim 0.4$ then  an AGB self-pollution
scenario, with a top-heavy IMF, might work. How then
to justify the existence of such an IMF for the first
generation of GC stars? There is some observational
evidence for variations in the IMF \citep{stolte05}
but there is  ample evidence supporting  a universal IMF,
at least in the field \citep{kroupa01}.
Moreover, none of the observational evidence for variations
in the IMF comes from environments similar to galactic GCs.
Dwarf spheroidal galaxies have a total mass comparable to the
largest clusters but supposedly did not have such strange 
IMFs \citep[see for e.g.][]{pritzl05}.
This may change as our ability to observe distant 
galaxies with young GCs increases, but it will
be a great challenge to extract a useful mass function 
from these systems.

\section{Conclusions}

Our investigation into the chemical evolution of helium in GCs 
highlights the difficulty the AGB self-pollution scenario suffers
in trying to explain the large postulated helium enrichment 
required to fit the horizontal branch of clusters like NGC 2808.
With a standard Salpeter IMF, the largest
predicted helium abundance in the cluster gas is $Y \approx 0.29$
but this is accompanied by a large increase in the C$+$N$+$O abundance.
Using an independent set of AGB yields from \citet{ventura02} we find a
maximum $Y \approx 0.26$ and only a modest increase of
C$+$N$+$O $\approx 0.4$~dex,  probably within the
observational errors.  
We conclude that with a standard IMF it does not seem likely that
the AGB self-pollution mechanism alone produced the enormous 
amounts of helium inferred from observations of the bluest HB and 
main-sequence stars of clusters like NGC 2808 and $\omega$ Centauri.

Simulations that employ the
IMS-enhanced IMF show larger helium enhancements  of
$Y \approx 0.35$ but only  when accompanied by
enormous increases in the total C$+$N$+$O content of the cluster
gas, in violation of observations. The Ventura et al. yields
predict a maximum $Y \approx 0.28$ and the total CNO abundance
stays constant to within $\sim 0.3$\,dex, although we again point
out that this maximum $Y$ is made uncertain by extrapolating the
yields to higher masses.
Even with such an extreme IMF we have a problem fitting
the observational constraints.  Indeed, the use of such
an IMF does not help the difficulties faced by the self-pollution
scenario in matching the constraints
that we have considered  in this study i.e. $Y \gtrsim 0.30$
and C$+$N$+$O $\approx$ constant. 
Both sets of yields considered here also fail to match the 
observed spread of abundances between O, Na, Mg and Al, even 
though the quantitative predictions of the models are different. 
For example, the Fenner et al. models produce too much Na whereas 
the Ventura et al. yields destroy sodium, resulting in a O-Na 
correlation.

\citet{bekki06} discussed the
consequences of a number of top-heavy IMFs on the evolution
of $\omega$ Centauri and concluded that the ones most suitable
for producing the large helium enrichment  would
also tear the cluster apart.  While $\omega$ Cen is an unusual
cluster indicated by the spread in Fe, and clearly had a very 
different chemical enrichment history to the other lower-mass GCs, 
the conclusions reached about the top-heavy IMF on the cluster
evolution are applicable to smaller mass, less tightly bound
systems. Given the difficulties associated with the self-pollution
scenario, we may need to look to other solutions such as pollution
from outside the cluster.  Indeed, perhaps the most unusual
of all clusters, $\omega$ Cen, is actually just an extreme member 
and is telling us something useful about all GCs.


%% Included in this acknowledgments section are examples of the
%% AASTeX hypertext markup commands. Use \url without the optional [HREF]
%% argument when you want to print the url directly in the text. Otherwise,
%% use either \url or \anchor, with the HREF as the first argument and the
%% text to be printed in the second.

\acknowledgments

\section*{Acknowledgments}

The authors wish to thank Chris Matzner, Debra Shepherd and Ralph Pudritz
for useful discussions. AIK wishes to thanks the H. G. Thode Foundation for 
financial support. This work was partially supported by the Australian
Research Council.

\clearpage


\centerline{\bf FIGURE CAPTIONS}

\figcaption[fig1]{\label{fig1} Choices of the IMF for the first
generation of GC stars.  We use a standard Salpeter IMF with slope
$s = 1.31$ and a IMS-enhanced IMF that places about 10 times
more mass in 3.5 to 6.5$\Msun$ stars.}

\figcaption[fig2]{\label{fig2}  The evolution of helium ($Y$) as a function 
of time (Gyr) in the cluster gas. Here we assume a standard Salpeter IMF 
and we show the predicted helium abundance using our yields (solid line) 
and yields from \citet{ventura02} (dashed line). In (a) we show the results 
with some dilution with primordial material and in (b) with no dilution (i.e. 
using pure AGB ejecta). }

\figcaption[fig3]{\label{fig3} The temporal evolution of the C$+$N$+$O 
abundance ([CNO/Fe]) in the cluster gas as a function of $Y$ for the Salpeter 
IMF and some dilution with primordial material, as discussed in the text.
We use the standard spectroscopic notation 
[X/Fe] = $\log (X/Fe) - \log(X/Fe)_{\odot}$, with solar abundances
from \cite{anders89}.  Results using our yields are shown in (a) and
results using Ventura et al. yields in (b).}

\figcaption[fig4]{\label{fig4} Same as Figure~\ref{fig3} but showing results for 
the IMS-enhanced IMF. }

\figcaption[fig5]{\label{fig5} The temporal evolution of the C$+$N$+$O species 
as a function of the helium mass fraction $Y$ using a Salpeter IMF and no dilution
with primordial material.}

\clearpage

\begin{figure}
\plotone{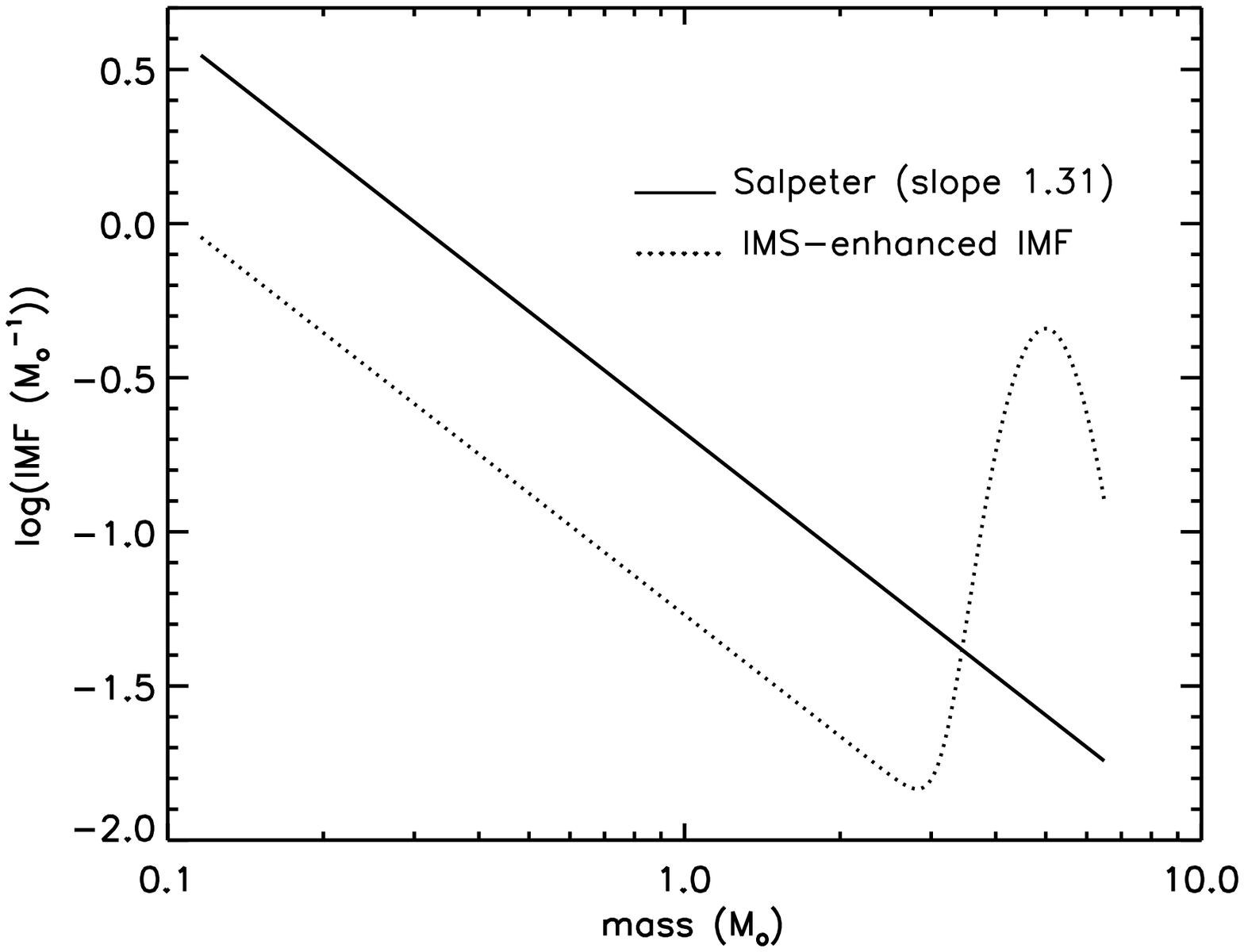}
\end{figure}

\clearpage

\begin{figure}
\plotone{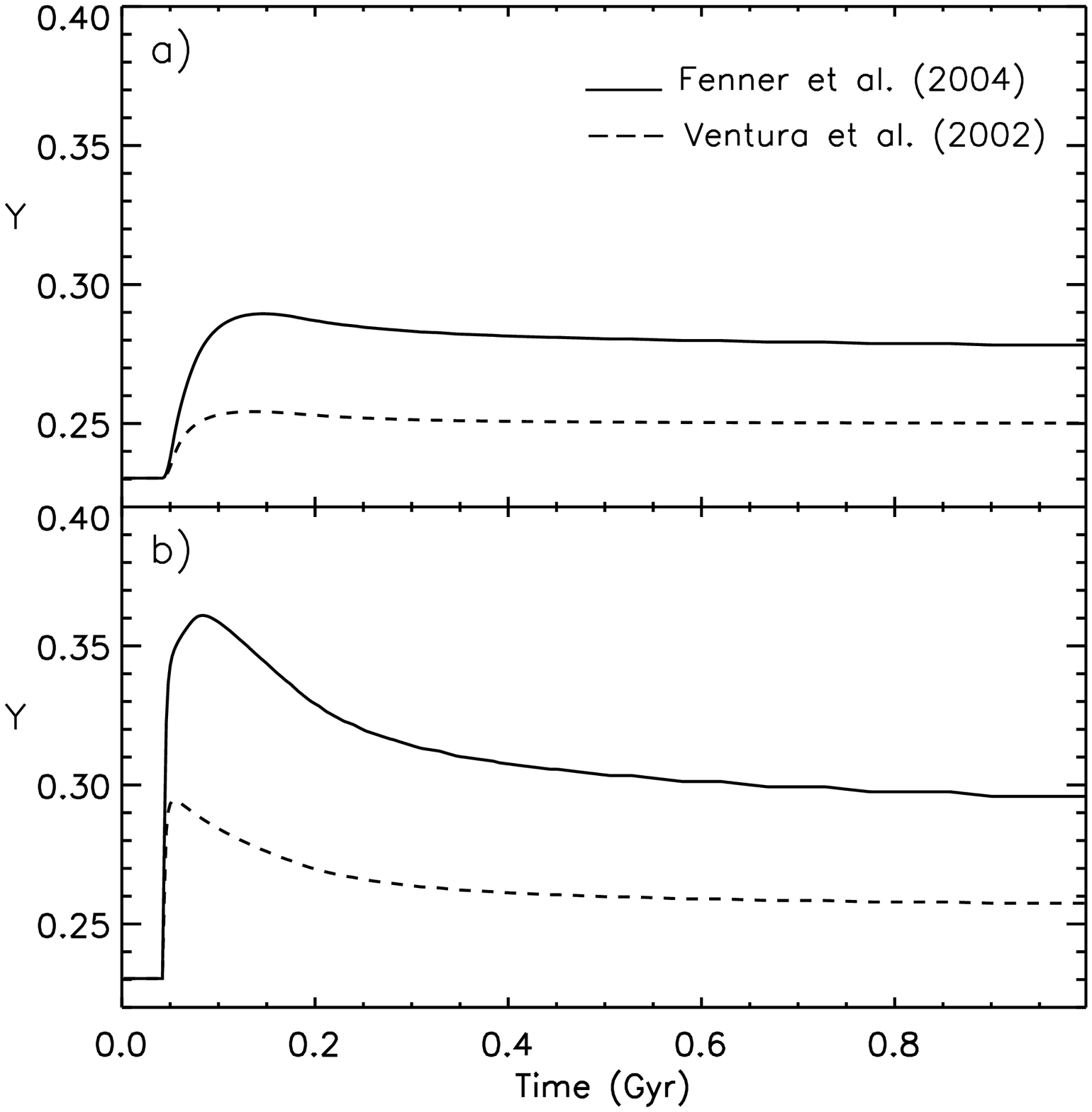}
\end{figure}

\clearpage

\begin{figure}
\epsscale{.9}
\plotone{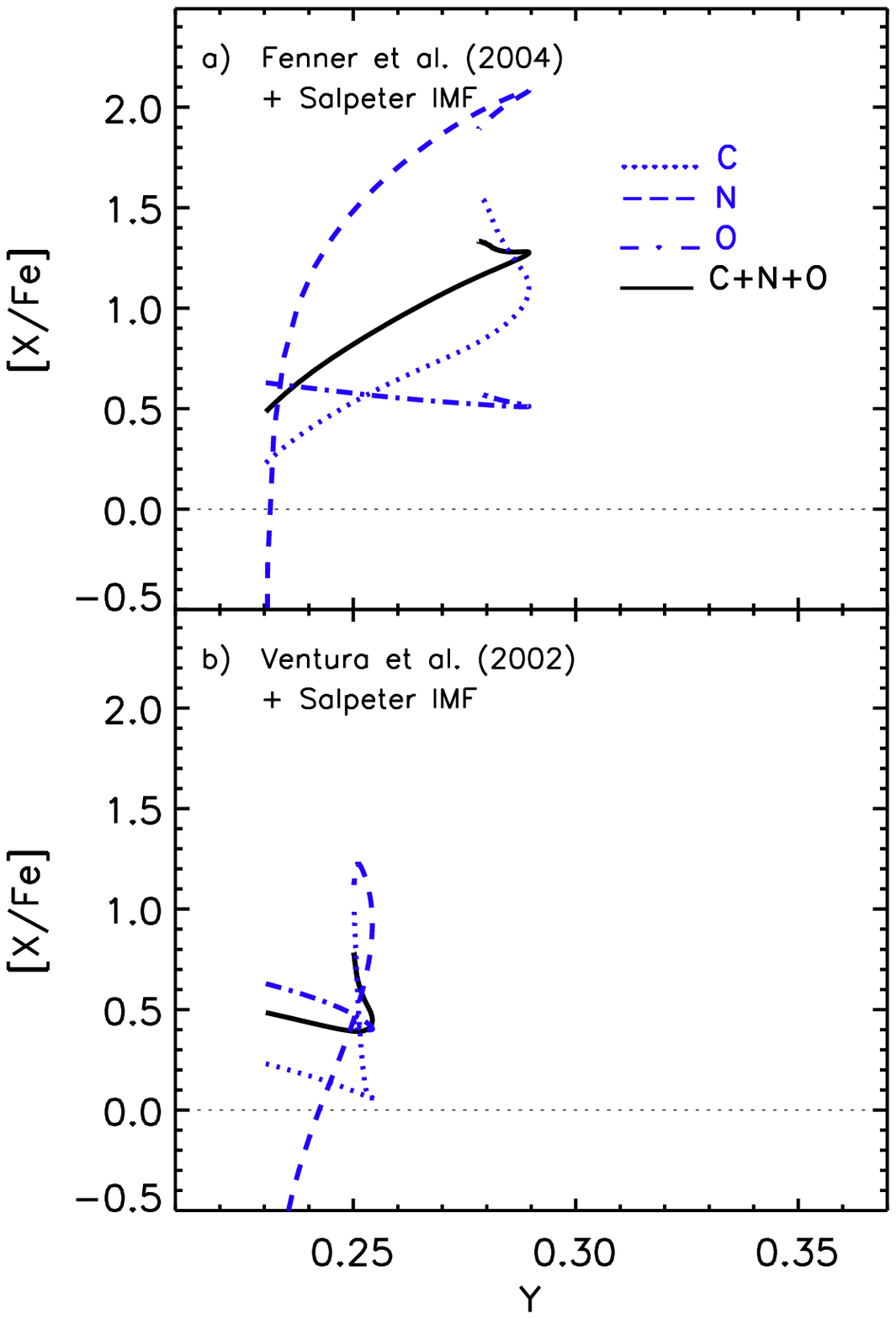}
\end{figure}

\clearpage

\begin{figure}
\plotone{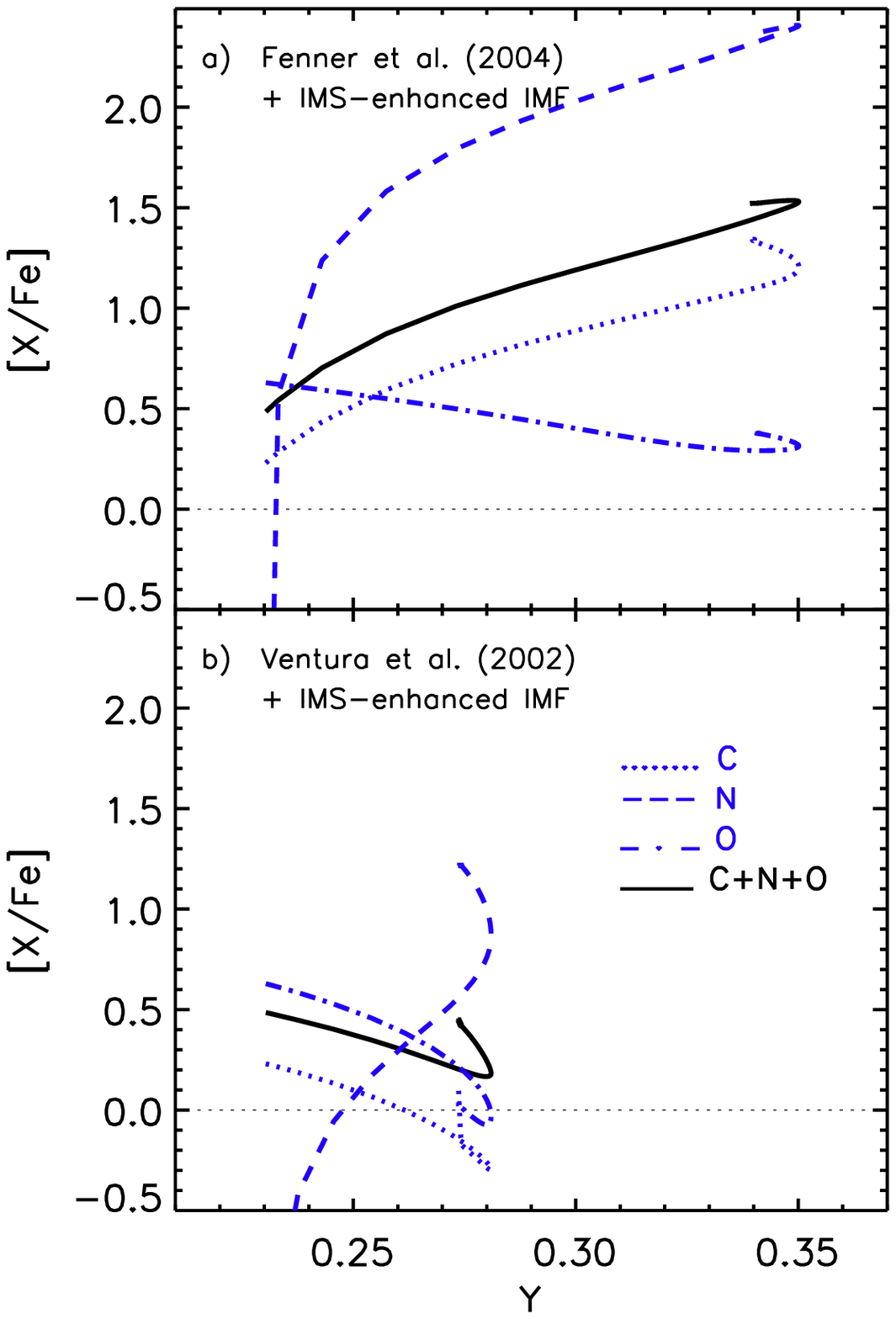}
\end{figure}

\clearpage

\begin{figure}%
\plotone{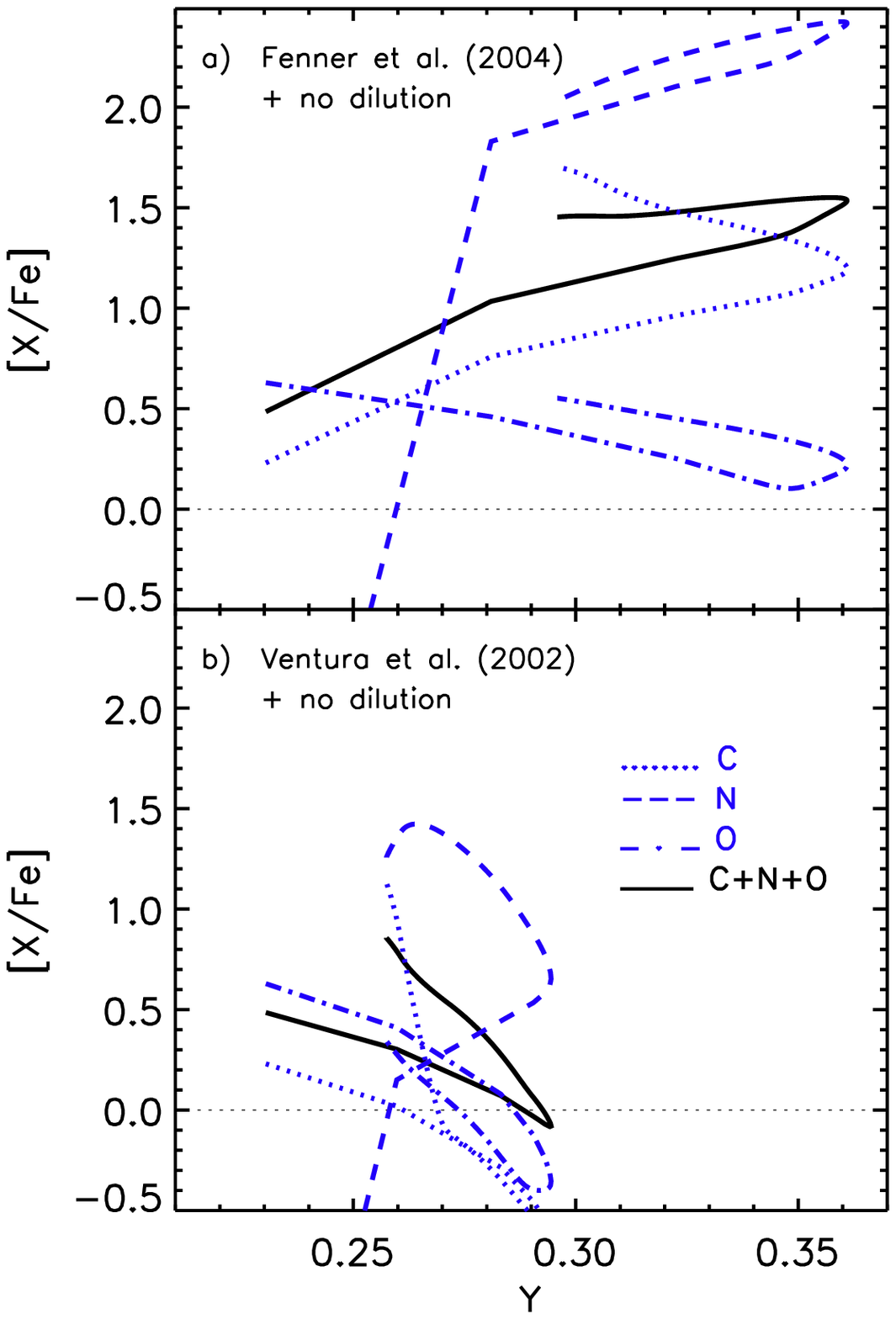}
\end{figure}

%% If you are not including electonic art with your submission, you may
%% mark up your captions using the \figcaption command. See the 
%% No more than seven \figcaption commands are allowed per page, 
%% so if you have more than seven captions, insert a \clearpage 

%% Tables should be submitted one per page, so put a \clearpage before

%% Two options are available to the author for producing tables:  the
%% deluxetable environment provided by the AASTeX package or the LaTeX
%% table environment.  Use of deluxetable is preferred.
%%

%% Three table samples follow, two marked up in the deluxetable environment,
%% one marked up as a LaTeX table.

%% In this first example, note that the \tabletypesize{}
%% command has been used to reduce the fonnt size of the table.
%% Note also that the \label command needs to be placed 

\clearpage

\begin{table}
\begin{center}
\caption{Yields of \iso{4}He expelled into the
intracluster medium.\label{table1}}
\begin{tabular}{lcccccc}
\tableline\tableline
Model & \multicolumn{6}{c}{Initial stellar mass ($\Msun$)} \\
      & & 2.5 & 3.5 & 5.0 & 5.5 & 6.5  \\
\tableline
\citet{campbell04} & $Y$  & 0.266 & 0.255 & 0.375 & -- & 0.349  \\
                   & mass & 0.490 & 0.680 & 1.52 & --  & 1.92  \\
\citet{ventura02}\tablenotemark{a} & $Y$ & --  & 0.257 & 0.289 & 0.293 &   \\
	       & mass & -- &  0.675 & 1.14 & 1.30 &  --  \\
\tableline
\end{tabular}

\tablenotetext{a}{\citet{ventura02} do not provide yields for 
masses less than 3$\Msun$, or for masses above 5.5$\Msun$.}

\end{center}
\end{table}


\begin{thebibliography}{57}
\expandafter\ifx\csname natexlab\endcsname\relax\def\natexlab#1{#1}\fi

\bibitem[{{Anders} \& {Grevesse}(1989)}]{anders89}
{Anders}, E. \& {Grevesse}, N. 1989, \gca, 53, 197

\bibitem[{{Bedin} {et~al.}(2000){Bedin}, {Piotto}, {Zoccali}, {Stetson},
  {Saviane}, {Cassisi}, \& {Bono}}]{bedin00}
{Bedin}, L.~R., {Piotto}, G., {Zoccali}, M., {Stetson}, P.~B., {Saviane}, I.,
  {Cassisi}, S., \& {Bono}, G. 2000, \aap, 363, 159

\bibitem[{{Bekki} \& {Norris}(2006)}]{bekki06}
{Bekki}, K. \& {Norris}, J.~E. 2006, \apjl, 637, L109

\bibitem[{{Boothroyd} \& {Sackmann}(1999)}]{boothroyd99}
{Boothroyd}, A.~I. \& {Sackmann}, I.-J. 1999, \apj, 510, 232

\bibitem[{{Campbell} {et~al.}(2004){Campbell}, {Fenner}, {Karakas},
  {Lattanzio}, \& {Gibson}}]{campbell04}
{Campbell}, S.~W., {Fenner}, Y., {Karakas}, A.~I., {Lattanzio}, J.~C., \&
  {Gibson}, B.~K. 2004, Memorie della Societa Astronomica Italiana, 75, 735

\bibitem[{{Cannon} {et~al.}(1998){Cannon}, {Croke}, {Bell}, {Hesser}, \&
  {Stathakis}}]{cannon98}
{Cannon}, R.~D., {Croke}, B.~F.~W., {Bell}, R.~A., {Hesser}, J.~E., \&
  {Stathakis}, R.~A. 1998, \mnras, 298, 601

\bibitem[{{Charbonnel}(1994)}]{charbonnel94}
{Charbonnel}, C. 1994, \aap, 282, 811

\bibitem[{{Chieffi} \& {Limongi}(2002)}]{chieffi02}
{Chieffi}, A. \& {Limongi}, M. 2002, \apj, 577, 281

\bibitem[{{Cohen} {et~al.}(2005){Cohen}, {Briley}, \& {Stetson}}]{cohen05b}
{Cohen}, J.~G., {Briley}, M.~M., \& {Stetson}, P.~B. 2005, \aj, 130, 1177

\bibitem[{{Cohen} \& {Mel{\'e}ndez}(2005)}]{cohen05a}
{Cohen}, J.~G. \& {Mel{\'e}ndez}, J. 2005, \aj, 129, 303

\bibitem[{{Cottrell} \& {Da Costa}(1981)}]{cottrell81}
{Cottrell}, P.~L. \& {Da Costa}, G.~S. 1981, \apjl, 245, L79

\bibitem[{{D'Antona} {et~al.}(2005){D'Antona}, {Bellazzini}, {Caloi}, {Pecci},
  {Galleti}, \& {Rood}}]{dantona05}
{D'Antona}, F., {Bellazzini}, M., {Caloi}, V., {Pecci}, F.~F., {Galleti}, S.,
  \& {Rood}, R.~T. 2005, \apj, 631, 868

\bibitem[{{D'Antona} \& {Caloi}(2004)}]{dantona04}
{D'Antona}, F. \& {Caloi}, V. 2004, \apj, 611, 871

\bibitem[{{Denissenkov} \& {Herwig}(2003)}]{herwig03b}
{Denissenkov}, P.~A. \& {Herwig}, F. 2003, \apjl, 590, L99

\bibitem[{{Fenner} {et~al.}(2004){Fenner}, {Campbell}, {Karakas}, {Lattanzio},
  \& {Gibson}}]{fenner04}
{Fenner}, Y., {Campbell}, S., {Karakas}, A.~I., {Lattanzio}, J.~C., \&
  {Gibson}, B.~K. 2004, \mnras, 353, 789

\bibitem[{{Forestini} \& {Charbonnel}(1997)}]{forestini97}
{Forestini}, M. \& {Charbonnel}, C. 1997, \aaps, 123, 241

\bibitem[{{Gratton} {et~al.}(2004){Gratton}, {Sneden}, \&
  {Carretta}}]{gratton04}
{Gratton}, R., {Sneden}, C., \& {Carretta}, E. 2004, \araa, 42, 385

\bibitem[{{Gratton} {et~al.}(2001){Gratton}, {Bonifacio}, {Bragaglia},
  {Carretta}, {Castellani}, {Centurion}, {Chieffi}, {Claudi}, {Clementini},
  {D'Antona}, {Desidera}, {Fran{\c c}ois}, {Grundahl}, {Lucatello}, {Molaro},
  {Pasquini}, {Sneden}, {Spite}, \& {Straniero}}]{gratton01}
{Gratton}, R.~G. et al., 2001, \aap, 369, 87

\bibitem[{{Gratton} {et~al.}(2000){Gratton}, {Sneden}, {Carretta}, \&
  {Bragaglia}}]{gratton00}
{Gratton}, R.~G., {Sneden}, C., {Carretta}, E., \& {Bragaglia}, A. 2000, \aap,
  354, 169

\bibitem[{{Harris}(1996)}]{harris96}
{Harris}, W.~E. 1996, \aj, 112, 1487

\bibitem[{{Izzard} {et~al.}(2004){Izzard}, {Tout}, {Karakas}, \&
  {Pols}}]{izzard04}
{Izzard}, R.~G., {Tout}, C.~A., {Karakas}, A.~I., \& {Pols}, O.~R. 2004,
  \mnras, 350, 407

\bibitem[{{James} {et~al.}(2004){James}, {Fran{\c c}ois}, {Bonifacio},
  {Bragaglia}, {Carretta}, {Centuri{\'o}n}, {Clementini}, {Desidera},
  {Gratton}, {Grundahl}, {Lucatello}, {Molaro}, {Pasquini}, {Sneden}, \&
  {Spite}}]{james04}
{James}, G. et al., 2004, \aap, 414, 1071

\bibitem[{{Karakas} {et~al.}(2006{\natexlab{a}}){Karakas}, {Fenner}, {Sills},
  {Campbell}, \& {Lattanzio}}]{karakas06b}
{Karakas}, A., {Fenner}, Y., {Sills}, A., {Campbell}, S., \& {Lattanzio}, J.
  2006{\natexlab{a}}, in The VIII Torino Workshop on Nucleosynthesis in AGB
  stars: Constraints on AGB Nucleosynthesis from Observations,
  astro--ph/0605540

\bibitem[{{Karakas} {et~al.}(2006{\natexlab{b}}){Karakas}, {Lugaro},
  {Wiescher}, {Goerres}, \& {Ugalde}}]{karakas06a}
{Karakas}, A., {Lugaro}, M., {Wiescher}, M., {Goerres}, J., \& {Ugalde}, C.
  2006{\natexlab{b}}, \apj, 643, 741

\bibitem[{{Karakas}(2003)}]{karakasThesis}
{Karakas}, A.~I. 2003, PhD thesis, Monash University

\bibitem[{{Kraft}(1994)}]{kraft94}
{Kraft}, R.~P. 1994, \pasp, 106, 553

\bibitem[{{Kraft} {et~al.}(1997){Kraft}, {Sneden}, {Smith}, {Shetrone},
  {Langer}, \& {Pilachowski}}]{kraft97}
{Kraft}, R.~P., {Sneden}, C., {Smith}, G.~H., {Shetrone}, M.~D., {Langer},
  G.~E., \& {Pilachowski}, C.~A. 1997, \aj, 113, 279

\bibitem[{{Kroupa}(2001)}]{kroupa01}
{Kroupa}, P. 2001, \mnras, 322, 231

\bibitem[{{Kroupa} {et~al.}(1993){Kroupa}, {Tout}, \& {Gilmore}}]{kroupa93}
{Kroupa}, P., {Tout}, C.~A., \& {Gilmore}, G. 1993, \mnras, 262, 545

\bibitem[{{Lattanzio} {et~al.}(2004){Lattanzio}, {Karakas}, {Campbell},
  {Elliott}, \& {Chieffi}}]{lattanzio04}
{Lattanzio}, J., {Karakas}, A., {Campbell}, S., {Elliott}, L., \& {Chieffi}, A.
  2004, Memorie della Societa Astronomica Italiana, 75, 322

\bibitem[{{Lee} {et~al.}(2005){Lee}, {Joo}, {Han}, {Chung}, {Ree}, {Sohn},
  {Kim}, {Yoon}, {Yi}, \& {Demarque}}]{lee05}
{Lee}, Y.-W. et al., 2005,
  \apjl, 621, L57

\bibitem[{{Lucatello} {et~al.}(2005){Lucatello}, {Gratton}, {Beers}, \&
  {Carretta}}]{lucatello05}
{Lucatello}, S., {Gratton}, R.~G., {Beers}, T.~C., \& {Carretta}, E. 2005,
  \apj, 625, 833

\bibitem[{{Marigo}(2001)}]{marigo01}
{Marigo}, P. 2001, \aap, 370, 194

\bibitem[{{Matzner} \& {McKee}(2000)}]{matzner00}
{Matzner}, C.~D. \& {McKee}, C.~F. 2000, \apj, 545, 364

\bibitem[{{Nakamura} \& {Umemura}(2001)}]{nakamura01}
{Nakamura}, F. \& {Umemura}, M. 2001, \apj, 548, 19

\bibitem[{{Norris}(2004)}]{norris04}
{Norris}, J.~E. 2004, \apjl, 612, L25

\bibitem[{{Pinsonneault}(1997)}]{pin97}
{Pinsonneault}, M. 1997, \araa, 35, 557

\bibitem[{{Piotto} {et~al.}(2005){Piotto}, {Villanova}, {Bedin}, {Gratton},
  {Cassisi}, {Momany}, {Recio-Blanco}, {Lucatello}, {Anderson}, {King},
  {Pietrinferni}, \& {Carraro}}]{piotto05}
{Piotto}, G. et al., 2005, \apj, 621, 777

\bibitem[{{Pritzl} {et~al.}(2005){Pritzl}, {Venn}, \& {Irwin}}]{pritzl05}
{Pritzl}, B.~J., {Venn}, K.~A., \& {Irwin}, M. 2005, \aj, 130, 2140

\bibitem[{{Ram{\'{\i}}rez} \& {Cohen}(2002)}]{ramirez02}
{Ram{\'{\i}}rez}, S.~V. \& {Cohen}, J.~G. 2002, \aj, 123, 3277

\bibitem[{{Ram{\'{\i}}rez} \& {Cohen}(2003)}]{ramirez03}
---. 2003, \aj, 125, 224

\bibitem[{{Salpeter}(1955)}]{salpeter}
{Salpeter}, E.~E. 1955, \apj, 121, 161

\bibitem[{{Shetrone}(1996)}]{shetrone96a}
{Shetrone}, M.~D. 1996, \aj, 112, 1517

\bibitem[{{Siess} {et~al.}(2000){Siess}, {Dufour}, \& {Forestini}}]{siess00}
{Siess}, L., {Dufour}, E., \& {Forestini}, M. 2000, \aap, 358, 593

\bibitem[{{Smith} {et~al.}(2000){Smith}, {Suntzeff}, {Cunha}, {Gallino},
  {Busso}, {Lambert}, \& {Straniero}}]{smith00}
{Smith}, V.~V., {Suntzeff}, N.~B., {Cunha}, K., {Gallino}, R., {Busso}, M.,
  {Lambert}, D.~L., \& {Straniero}, O. 2000, \aj, 119, 1239

\bibitem[{{Stolte} {et~al.}(2005){Stolte}, {Brandner}, {Grebel}, {Lenzen}, \&
  {Lagrange}}]{stolte05}
{Stolte}, A., {Brandner}, W., {Grebel}, E.~K., {Lenzen}, R., \& {Lagrange},
  A.-M. 2005, \apjl, 628, L113

\bibitem[{{Sweigart} \& {Mengel}(1979)}]{sweigart79}
{Sweigart}, A.~V. \& {Mengel}, J.~G. 1979, \apj, 229, 624

\bibitem[{{Thoul} {et~al.}(2002){Thoul}, {Jorissen}, {Goriely}, {Jehin},
  {Magain}, {Noels}, \& {Parmentier}}]{thoul02}
{Thoul}, A., {Jorissen}, A., {Goriely}, S., {Jehin}, E., {Magain}, P., {Noels},
  A., \& {Parmentier}, G. 2002, \aap, 383, 491

\bibitem[{{Umeda} \& {Nomoto}(2002)}]{umeda02}
{Umeda}, H. \& {Nomoto}, K. 2002, \apj, 565, 385

\bibitem[{{van den Hoek} \& {Groenewegen}(1997)}]{vandenhoek97}
{van den Hoek}, L.~B. \& {Groenewegen}, M.~A.~T. 1997, \aaps, 123, 305

\bibitem[{{Vassiliadis} \& {Wood}(1993)}]{vw93}
{Vassiliadis}, E. \& {Wood}, P.~R. 1993, \apj, 413, 641

\bibitem[{{Ventura} \& {D'Antona}(2005{\natexlab{a}})}]{ventura05a}
{Ventura}, P. \& {D'Antona}, F. 2005{\natexlab{a}}, \aap, 431, 279

\bibitem[{{Ventura} \& {D'Antona}(2005{\natexlab{b}})}]{ventura05b}
---. 2005{\natexlab{b}}, \aap, 439, 1075

\bibitem[{{Ventura} {et~al.}(2002){Ventura}, {D'Antona}, \&
  {Mazzitelli}}]{ventura02}
{Ventura}, P., {D'Antona}, F., \& {Mazzitelli}, I. 2002, \aap, 393, 215

\bibitem[{{White} \& {Hillenbrand}(2004)}]{white04}
{White}, R.~J. \& {Hillenbrand}, L.~A. 2004, \apj, 616, 998

\bibitem[{{Wylie} {et~al.}(2006){Wylie}, {Cottrell}, {Sneden}, \&
  {Lattanzio}}]{wylie06}
{Wylie}, E.~C., {Cottrell}, P.~L., {Sneden}, C.~A., \& {Lattanzio}, J.~C. 2006,
  ApJ, accepted, astro-ph/0605538

\bibitem[{{Yong} {et~al.}(2006){Yong}, {Aoki}, {Lambert}, \&
  {Paulson}}]{yong06b}
{Yong}, D., {Aoki}, W., {Lambert}, D.~L., \& {Paulson}, D.~B. 2005, ApJ,
  639, 918

\end{thebibliography}
\end{document}